# Hybrid bound states in the continuum in terahertz metasurfaces

Junxing Fan[1], Zhanqiang Xue[1], Hongyang Xing[1], Dan Lu[1], Guizhen Xu[1], Jianqiang Gu[2,4], Jiaguang Han[2,3,5] and Longqing Cong[1,*]

[1]Department of Electrical and Electronic Engineering, Southern University of Science and Technology, Shenzhen 518055, China

[2]Center for Terahertz Waves and College of Precision Instrument and Optoelectronics Engineering, Tianjin University, Tianjin 300072, China

[3]Guangxi Key Laboratory of Optoelectronic Information Processing, School of Optoelectronic Engineering, Guilin University of Electronic Technology, Guilin 541004, China

[4]gjq@tju.edu.cn
[5]jiaghan@tju.edu.cn
[*]conglq@sustech.edu.cn

**Abstract:** Bound states in the continuum (BICs) have exhibited extraordinary properties in photonics for enhanced light-matter interactions that enable appealing applications in nonlinear optics, biosensors, and ultrafast optical switches. The most common strategy to apply BICs in a metasurface is by breaking symmetry of resonators in the uniform array that leaks the otherwise uncoupled mode to free space and exhibits an inverse quadratic relationship between quality factor ($Q$) and asymmetry. Here, we propose a scheme to further reduce scattering losses and improve the robustness of symmetry-protected BICs by decreasing the radiation density with a hybrid BIC lattice. We observe significant increase of radiative $Q$ in the hybrid lattice compared to uniform lattice with a factor larger than 14.6. In the hybrid BIC lattice, modes are transferred to Γ point inherited from high symmetric X, Y and M points in the Brillouin zone that reveal as multiple Fano resonances in the far field and would find applications in hyperspectral sensing. This work initiates a novel and generalized path toward reducing scattering losses and improving the robustness of BICs in terms of lattice engineering that would release the rigid requirements of fabrication accuracy and benefit applications of photonics and optoelectronic devices.

**Introduction**

Resonant cavities can effectively confine light and enhance light-matter interactions, which are of great importance to technologies and applications including lasers[1-3], filters[4, 5], harmonic generation[6-10], and sensors[11, 12]. Quality factor ($Q$) is one of the most important parameters to characterize the strength of light-matter interactions. Different schemes have been proposed to improve $Q$ in microcavities[13], such as microdisks[14, 15], Bragg reflector microcavities[16], and photonic crystals[17]. In the same context, a generalized concept of bound states in the continuum (BIC) was raised[18, 19], which is in fact a topological defect in the momentum space and localizes in the continuous spectrum of extended states but is unable to couple to free space. Therefore, an ideal BIC will theoretically exhibit an infinite lifetime. Since the first demonstration, BICs have been applied to improve the performance in a plethora of optical applications by opening a coupling channel to free space via symmetry breaking of unit cells. Although any value of $Q$ could be theoretically obtained, the measured values are commonly much lower than their theoretical predictions due to unavoidable scattering losses by fabrication defects, a finite size of resonator array, and Ohmic losses in practice[20-22]. One solution to reduce the susceptibility of BIC to defect-induced scattering losses is by merging multiple BICs in the vicinity of Γ point in the Brillouin zone (BZ) that in fact decreases the slope of radiative $Q$ versus wavevector ($k$) from an inverse quadratic relationship ($Q_{rad} \propto k^{-2}$) to a higher order (e.g., $Q_{rad} \propto k^{-6}$) and thus improves the rubustness[23-25]. This is a smart strategy to access a stable and high $Q$ resonance; however, a large number of complex calculations and sub-nanometer geometrical accuracy of resonators are necessary to guarantee the merged states, and a special band is demanded that must possess multiple BICs at Γ point as well as off-Γ points (accidental BICs).

Here, we introduce a generalized scheme to access robust and high-$Q$ BICs by decreasing the radiation density in a *hybrid* metasurface lattice. For a common lattice supporting symmetry-protected BICs, the leaky channel is opened by uniformly breaking the symmetry of resonators, while a *hybrid lattice* indicates that half or a quarter (or less) of resonators in the array are leaky so that the radiation density reduces. In the course of radiation suppression, the radiative $Q$ versus $k$ will theoretically have a 16-time larger coefficient in a 1/4 hybrid lattice compared to the common BICs that would enable a robust high-$Q$ resonance. The idea was numerically and experimentally demonstrated by an array of classical double-gap split ring resonators (DSRR) with $C_2$ symmetry in terahertz regime that supports symmetry-protected BICs[26]. Band folding analysis in BZ was performed to visualize the evolution of modes and quality factors between *uniform* and *hybrid* lattices. In the hybrid lattice, the high-$Q$ portion of band was retained from the uniform lattice, while the low-$Q$ portion was discarded in the band folding process resulting in a relatively high $Q$ (more than 14.6 times higher than uniform lattice in simulations) and robust BIC. Accompanying with the band folding, certain inaccessible modes in uniform lattice become accessible at Γ point, and multiple Fano resonances were captured in the hybrid lattice which could enable an alternative approach for broadband molecular fingerprint sensing[11, 12].

**Results and discussion**

We demonstrate the idea with array of classical DSRRs which could be generalized to other symmetry-protected BIC scenarios without prerequisite of multiple BICs in the band. In the symmetric scenario of DSRR, symmetry-protected BIC is uncoupled to free space exhibiting an infinite radiative $Q$ as illustrated in Fig.1a. A common strategy to observe the mode in the far field is to break the symmetry of DSRRs by displacing one gap from the center, and radiative channel of the mode is thus open defined as a quasi-BIC (qBIC)[27]. When the unit cells in the metasurface are uniformly displaced (Fig. 1b), the well-understood type of BIC is defined as a uniform quasi-BIC (U-qBIC) whose radiative $Q$ depends on $\alpha$ following the inverse quadratic law[28, 29]

$$Q_{rad} = m\alpha^{-2} \qquad (1)$$

where $\alpha$ is the asymmetric degree (AD) of DSRR and $m$ is a constant determined by geometric parameters. The AD of DSRR is defined as $\alpha = (l_1 - l_2)/(l_1 + l_2) \times 100\%$, where $l_1$ and $l_2$ denote the total lengths of left and right metallic branches of the resonator, respectively (see Fig. 1d).

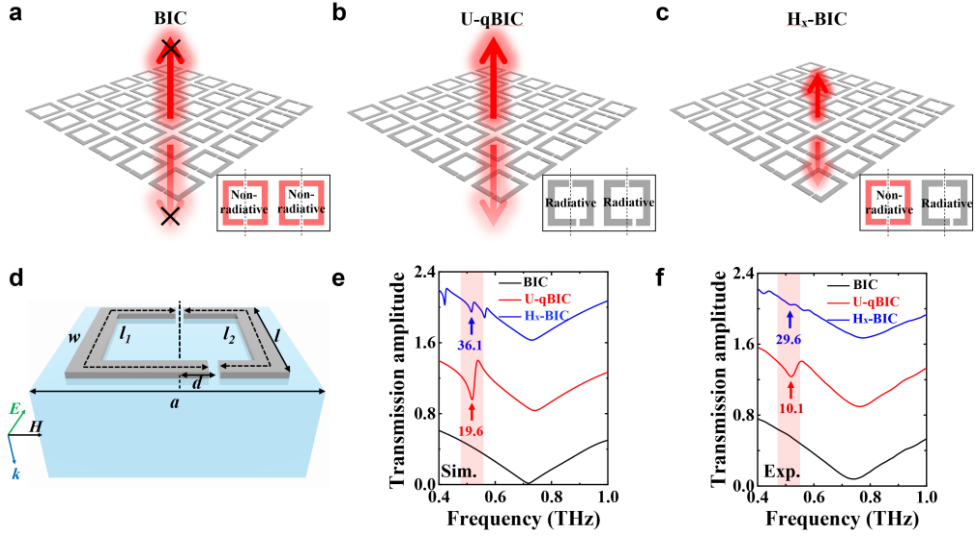

**Fig. 1| Hybrid BIC lattices.** (**a-c**) Schematic diagram of symmetry-protected BIC lattice without radiation channel (a), uniform quasi-BIC lattice with radiation channel open for all the resonators by breaking symmetry (b), and hybrid quasi-BIC lattice with half radiation channel open interchanging along $x$ axis (c). (**d**) A double gap split ring resonator as the unit cell of the metasurface. (**e, f**) Simulated (e) and experimental (f) transmission amplitude spectra for the three-type lattices. The same asymmetry degree ($\alpha = 4.95\%$) was applied for U-qBIC and $H_x$-BIC metasurfaces.

In the scenario of U-qBIC, all the symmetry-breaking resonators contribute to the far-field radiation that follows the inverse quadratic law in Eq. 1. In addition to the general strategy of breaking symmetry to induce U-qBIC, we propose a hybrid BIC ($H_x$-BIC, Fig. 1c) metamolecule to reduce the radiative loss: close the radiative channel of neighboring resonators along $x$ axis by recovering them back to $C_2$ symmetric. When we set the sizes of resonators with $l = 60$ μm, width $w = 8$ μm, gap $g = 3$ μm, and period of square lattice $a = 73$ μm (Fig. 1d), the frequency of fundamental BIC will fall in terahertz regime. Simulated and experimental far-field transmission spectra of the three scenarios are shown in Fig. 1e and 1f, respectively (at $\alpha = 4.95\%$ for U-qBIC and $H_x$-BIC). By comparing BIC, U-qBIC and $H_x$-BIC in the spectra, we observe the evolution of quasi-BIC as revealed by the Fano resonances in the far field where the resonance cannot be captured for symmetric resonators ($\alpha = 0$, black line), and appears at 0.52 THz in DSRR with $\alpha = 4.95\%$ (red and blue lines) for U-qBIC and $H_x$-BIC lattices. Quality factors of the Fano resonances were extracted by using Fano fitting (see Methods). A significant improvement of $Q$ is observed from U-qBIC ($Q$ = 10.1) to $H_x$-BIC ($Q$ = 29.6) in the experiments (Fig. 1f, from 19.6 to 36.1 in simulations as shown in Fig. 1e).

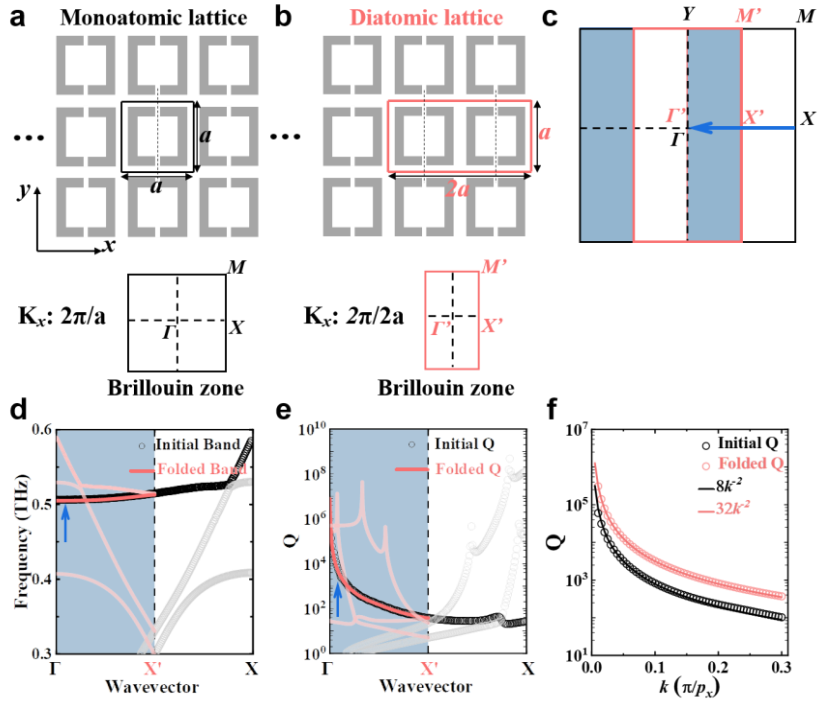

**Fig. 2| Interpretation of hybrid BIC from reciprocal space.** (**a, b**) Brillouin zones of monoatomic and diatomic lattices at $\alpha=0$ when the periods were chosen with $a$ and $2a$ in the $x$ direction. (**c**) Illustration of Brillouin zones for monoatomic and diatomic lattices showing the band folding operation where X and M points in the BZ of monoatomic lattice are folded to X' and M' points in the BZ of diatomic lattice, and $\Gamma(\Gamma')$ point is fixed. (**d**) Band diagrams of monoatomic (black circles) and diatomic (orange lines) lattices showing the folding behavior where all the modes of monoatomic in X' to X are reflected into the shaded region of diatomic with respect to the dashed line. (**e**) Radiative quality factors of monoatomic and

diatomic lattices. The same folding behavior of $Q$ is inherited from the eigenmodes. (**f**) Comparison of radiative $Q$ between monoatomic and diatomic lattices under the same wavevector. Circles are simulated values and solid lines are fitting curves with Eq. 1 whose coefficient is 4-times in diatomic larger than that of monoatomic. Here, $p_x$ is lattice period along $x$ direction. Perfect electric conductor (PEC) was used for DSRRs in simulations to calculate eigenvalues and quality factors.

The intuitive evolution of BIC from U-qBIC to $H_x$-BIC is unambiguously interpreted from the eigenvalue analysis in the reciprocal space. As shown in Fig. 2a and 2b, the period of diatomic lattice is doubled along $x$ axis compared with monoatomic one (with $\alpha=0$ to obtain the intrinsic properties), and the corresponding BZ is thus folded accordingly in Γ-X direction as indicated in Fig. 2c. The outer black square and inner orange rectangle describe the BZ of monoatomic and diatomic lattices, respectively. The energy dispersion surfaces have inversion symmetry in BZ, and thus the eigenvalues in blue and white-colored areas are equivalent as determined by the time-reversal symmetry[30]. Such a unique property leads to a folding BZ of diatomic lattice from monoatomic lattice, and modes at the edge of the unfolded BZ (X and M points) in monoatomic lattice are folded to Γ and Y points[29, 31, 32].

The mode evolution in the band folding process was verified with eigenmode analysis in simulations with DSRRs at $\alpha=0$ to obtain the eigenfrequencies and intrinsic radiative $Q$ of the interested modes (see Methods). In Fig. 2d, the full band diagram of diatomic lattice was shrunk in $k$ axis (indicated by Γ-X', orange lines) to half of monoatomic lattice (Γ-X, black dots) for comparison where a coincidence of the BIC band (highlighted by an arrow) occurs but with only half band. All the remaining half bands (from X'-X) of monoatomic lattice appear in diatomic lattice as reflection with respect to the dotted line located at $k=\pi/2a$. The direct consequence of the band folding process is that the otherwise inaccessible modes at X point in monoatomic lattice become leaky at Γ point of diatomic lattice which could thus be observed as multiple Fano resonances in the far field (as revealed by the Fano resonances in blue lines of Fig. 1e and 1f). The accompanying radiative $Q$ of the modes inherit the same folding properties in BZ (Fig. 2e). Interestingly, the band folding process discards the half BIC band with lower $Q$ at larger wavevectors in monoatomic (folded as a new band in the blue BZ) whose radiative $Q$ and $k$ relationship still follows Eq. 1. The half band with higher $Q$ in monoatomic lattice is expanded to fill the full BZ of diatomic lattice, and thus will reveal a larger and more stable quality factor. A direct comparison of $Q$ in the same wavevector is shown in Fig. 2f where the radiative $Q$ of diatomic lattice is always larger than that of monoatomic, and the numerical fitting with Eq. 1 reveals a 4-times increase in coefficient as a consequence of the folding and expanding process. The 4-times higher radiative $Q$ originates from the fact that half of resonators cease to radiate in $H_x$-BIC lattice which is inversely squared to contribute to radiative quality factors.

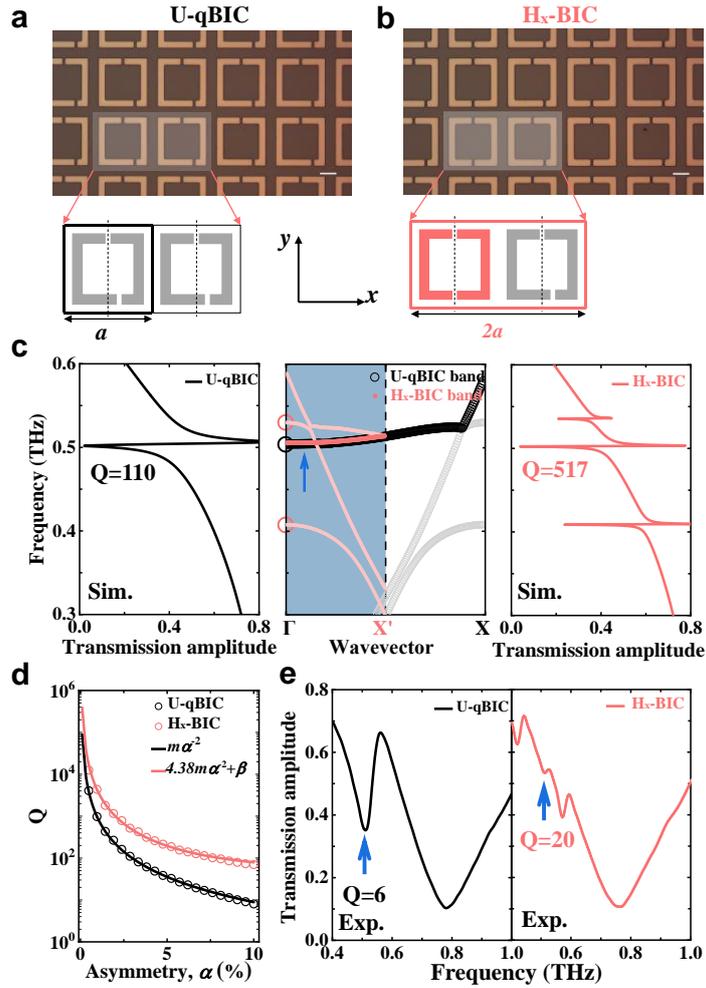

**Fig. 3| Experimental demonstration of stable and high-$Q$ hybrid BIC.** (**a, b**) Microscopic images of U-qBIC and H$_x$-BIC metasurfaces. The unit cells of U-qBIC and H$_x$-BIC metasurfaces are shown in the inset. Scale bar, 20 μm. (**c**) The simulated transmission amplitude spectra of U-qBIC (left) and H$_x$-BIC (right) metasurfaces at asymmetry degree of 2.97% with excitation electric field polarized along $y$ axis. The band structures of U-qBIC and H$_x$-BIC metasurfaces are shown in the middle. (**d**) Simulated radiative $Q$ (circles) versus asymmetry degree ($\alpha$) with inverse quadratic fitting curves (solid lines) of U-qBIC (black) and H$_x$-BIC (orange) metasurfaces. Here, an additional constant of $\beta$ is necessary to account for the nonuniform asymmetry in the hybrid lattice. (**e**) Experimental transmission amplitude spectra of U-qBIC and H$_x$-BIC metasurfaces at asymmetry degree of 7.42% with excitation electric field polarized along $y$ axis. The linewidth of Fano resonances is larger than that of simulations due to Ohmic loss in metallic resonators (aluminum) and finite number of excited resonators.

The above discussion could be numerically and experimentally demonstrated via far-field measurements in transmission spectra by breaking symmetry of the resonators (i.e., $\alpha \neq 0$) which would shift BIC to quasi-BIC at Γ point (Fig. 3a and 3b). All the resonators are asymmetric in U-qBIC (monoatomic) lattice with a period of $a$, and H$_x$-BIC (diatomic) lattice is constructed by restoring C$_2$ symmetry of the neighboring resonators in the diatomic unit cells so that the period along $x$ direction is $2a$ (Fig. 3b). Far-field transmission spectra were calculated with normal incidence (along $z$ direction) using COMSOL Multiphysics (see methods), and the leaky quasi-BIC was captured as a classical Fano line shape in Fig. 3c for U-qBIC lattice (at $\alpha=2.97\%$). For H$_x$-BIC lattice, two more Fano resonances appear in addition to the original one as a consequence of band folding that originate from X point of U-qBIC lattice (right panel of Fig. 3c). All the three Fano resonances match with the mode frequencies in the folded band at Γ point (orange lines in the middle panel of Fig. 3c). Although the band dispersion coincides between U-qBIC and H$_x$-BIC lattices for the quasi-BIC mode, their radiative $Q$ versus $k$ relationship reveals a divergence (supplementary Fig. S1), and the fitting curves still follow the inverse quadratic law but with different coefficients due to the involvement of geometrical asymmetry. The band folding and expansion processes lead to a significant improvement of $Q$ of Fano resonances at 0.504 THz from $Q_{rad} = 110$ (U-qBIC) to $Q_{rad} = 517$ ($H_x$-BIC). For asymmetric resonators, radiative $Q$ versus $α$ relationship reveals similar inverse quadratic dependence as that of radiative $Q$ versus $k$, and significant improvement of $Q$ values and larger robustness against variation of $α$ are obtained in $H_x$-BIC lattice (Fig. 3d).

We fabricated the samples with conventional photolithography, and the microscopic images of samples are shown in Fig. 3a and 3b. The transmission spectra were measured with terahertz time-domain spectroscopy (THz-TDs) system. Typical spectral features of Fano resonances were captured in experiments except the overall lower quality factors and weaker resonant features (Fig. 3e) that are attributed to the intrinsic losses of metallic resonators, finite number of excited resonators, limited scan length (40 ps) and signal to noise ratio of the experimental setup. We designed the samples at a larger asymmetry degree of 7.42% so that the quality factor falls within the instrumental resolution. The same interpretation could be simply generalized to $H_y$-BIC lattice which folds the band of Γ-Y (see supplementary Fig. S2 with detailed band analysis, simulations and experiments) and $H_d$-BIC lattice which folds the band of Γ-M (see supplementary Fig. S3).

With the interpretation of band folding in the hybrid lattice, we could further expand the scheme to high order in the BZ by introducing a diagonal non-radiative resonator in the 2×2 square lattice. In this scenario, two configurations, where one (Fig. 4a, $H_q$-BIC) and three (Fig. 4b, $H_t$-BIC) out of the four resonators in a supercell are left to radiate, will share exactly the same band diagram (middle panel of Fig. 4d) folded from that of U-qBIC lattice. However, their $Q$ will exhibit a striking divergence as a result of different radiation densities (see supplementary Fig. S4). The calculated band diagram and transmission spectra are shown in Fig. 4d where seven Fano resonances are observed for both $H_q$-BIC and $H_t$-BIC lattices at $α = 2.97\%$.

Fano frequencies match with mode frequencies at Γ point in the reciprocal space where all the modes are folded from X (red circles), Y (orange circles), and M (blue circles) points except the original black line. Despite of the same band diagram, a significantly narrower linewidth is observed in the original qBIC mode at 0.504 THz for $H_q$-BIC due to the less radiative channels than that of $H_t$-BIC (highlighted resonances in Fig. 4d). It is noted that the significant improvement of radiative $Q$ in $H_q$-BIC occurs only in the vicinity of Γ, and gradually converges to overlap at off-Γ points in reciprocal space since all the resonators gradually increase radiation at large wavevectors (see supplementary Fig. S4c for detailed analysis). Samples with $H_q$-BIC and $H_t$-BIC lattices at $\alpha = 7.42\%$ (Fig. 4a and 4b) were fabricated and measured with far-field transmission spectra as shown in Fig. 4e. A larger asymmetry degree was applied in the samples for measurements since the limited resolution and signal to noise ratio of THz-TDs led to the difficulty in capturing the very high-$Q$ resonances. All the seven Fano resonances can be captured but reveal weak spectral signatures as a result of finite number of excited unit cells with inevitable Ohmic losses as compared to the infinite array in simulations.

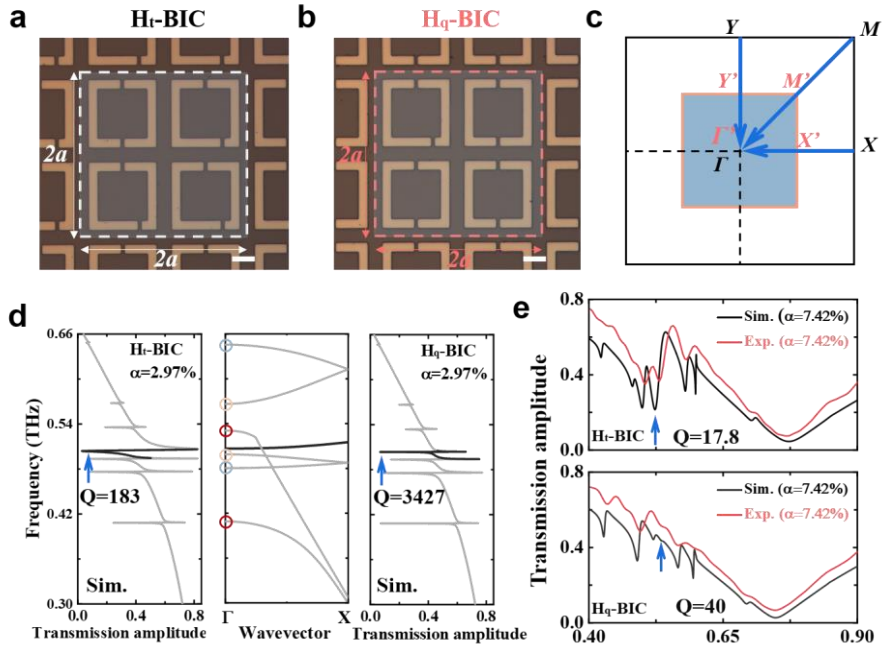

**Fig. 4| Generalized high-order hybrid BIC.** (**a, b**) Microscopic images of $H_t$-BIC and $H_q$-BIC metasurfaces with three and one asymmetric resonators out of four in a 2×2 supercell,

respectively, and the period is 2*a* along both *x* and *y* axes. Scale bar, 20 μm. (**c**) Schematic diagram of band folding from U-qBIC lattice (black) to H$_t$-BIC/H$_q$-BIC (red) in the Brillouin zone. (**d**) Simulated transmission amplitude spectra of the H$_t$-BIC (left) and H$_q$-BIC (right) metasurfaces at asymmetry degree of 2.97%. The band structure of H$_t$-BIC/H$_q$-BIC is shown in the middle, and the modes at the Γ point marked with different colored circles are folded from X (red), Y (orange) and M (blue) points in the Brillouin zone of U-qBIC lattice, respectively. The highlighted resonances show the original modes inherited from U-qBIC lattice. (**e**) Experimental (orange) and simulated (black) transmission amplitude spectra of H$_t$-BIC and H$_q$-BIC metasurfaces at asymmetry degree of 7.42%. The overall linewidth of Fano resonances is larger than that of simulations due to Ohmic loss in metallic resonators (aluminum) and finite number of excited resonators.

Symmetry-protected BICs have found a plethora of important applications in lasing, nonlinear optics, terahertz generation, and biosensors. Significant improvement of light-matter interactions could be obtained with these BICs that have enabled lower lasing threshold, higher-efficiency generation of harmonics and terahertz radiations, and hyperspectral sensing. Traditionally, most common logic to access quasi-BIC is to break the symmetry in the level of resonator itself. In this work, we have introduced a different approach to tailor the radiative losses from the level of the entire lattice. Via selectively preserving the C$_2$ symmetry of the sub-cell in a hybrid BIC lattice, we can reduce the radiation density and thus effectively improve the overall quality factors of the modes. We numerically and experimentally investigated four-type typical hybrid lattices via ceasing 0/4 (U-qBIC), 1/4 (H$_t$-BIC), 2/4 (H$_{x/y/d}$-BIC), and 3/4 (H$_q$-BIC) radiative channels in the lattice, and found a progressive increase of the overall quality factors (Fig. 5a). The retrieved $Q_{rad}$ from experiments reveals a good agreement with simulations and theory. An amplification coefficient of 14.6 of $Q_{rad}$ in H$_q$-BIC lattice ($Q = 59363$) is observed compared with that in U-qBIC lattice ($Q = 4062$) at a fixed asymmetry degree of $\alpha = 0.495\%$ (16 times for intrinsic BIC from theory of band folding), and a larger contrast is observed at larger asymmetry degrees. In addition, H$_q$-BIC reveals a clue of saturation of radiative $Q$ at larger asymmetries which guarantees that it will not deteriorate due to the fabrication imperfection or disorders.

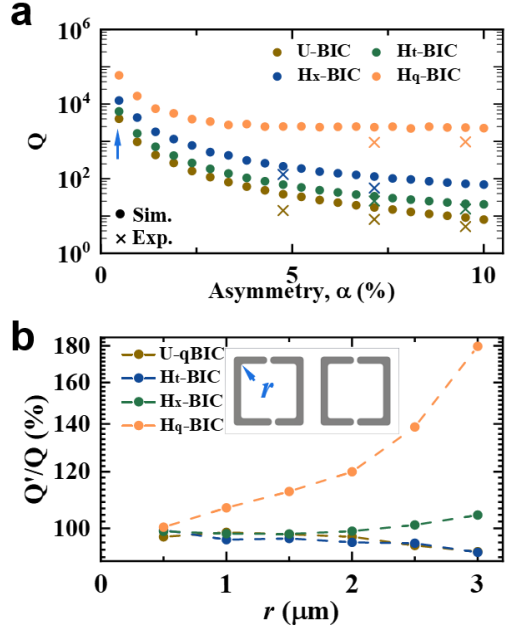

**Fig. 5| Significant *Q* improvement in hybrid BIC lattices and robustness against fabrication imperfections.** (**a**) Evolution of radiative *Q* versus asymmetry degree for U-qBIC, H$_t$-BIC, H$_x$-BIC and H$_q$-BIC lattices. The overall quality factors are improved in hybrid unit cells with a lower radiation density. (**b**) Influences of fabrication imperfection on quality factors in the four scenarios. The imperfection is introduced by adjusting the sharp right-angle square in the resonators to rounded angles indicated by radius *r*. *Q* and *Q*' indicate the radiative quality factors for lattices with right-angle and rounded angle resonators, respectively.

A common fabrication imperfection in photolithography and E-beam lithography is the rounded angles in the square or rectangular resonators instead of the designed right angles. The radius (*r*) of rounded angles is usually impossible to be accurately predicted which will thus deteriorate the very sensitive high-*Q* resonances as shown in Fig. 5b for U-qBIC and H$_t$-BIC lattices. However, the robustness of quality factors is better in H$_x$-BIC and H$_q$-BIC lattices that will guarantee high quality factors to a large extent. It is noted that the saturation of radiative *Q* in H$_q$-BIC lattice enables an obvious increase of quality factors at a larger defect radius. Simulated transmission spectra of samples with different rounded angles are shown in supplementary Fig. S5.

**Discussion**

In summary, we report a generalized strategy to improve quality factors of symmetry-protected BICs with hybrid BIC lattices. The idea is interpreted from reciprocal space, and the stable and higher quality factors of hybrid BIC lattices stem from band folding operation in the folded Brillouin zone. The consequent multiple Fano resonances at Γ point due to band folding provide an excellent solution for hyperspectral sensing and would be useful for high quality applications in optoelectronic devices. The hybrid BIC lattices exhibit great robustness to fabrication imperfections and disorders and would release the rigid requirements of fabrication accuracy, especially for applications requiring extremely precise quality factors.

**Methods**

**Simulation**

Numerical simulations were carried out using commercially available software (COMSOL Multiphysics) with RF module of the finite-element frequency-domain solver. The periodic boundary conditions were employed for the square lattice. Perfectly matching layer (PML) was applied at the input and output ports. A constant refractive index of $n = 3.45$ was set for silicon without loss as substrate. For the calculations of eigenvalues and radiative quality factors, perfect electric conductor (PEC) was used for the metallic part of DSRR resonators in simulations. In simulation of transmission spectra, metallic part of DSRR was set as aluminum with a conductivity of $3.56 \times 10^7$ S·m$^{-1}$.

**Fabrication**

Prior to fabrication, a 500 μm thick high-resistivity silicon (resistivity > 10000 Ω·cm) wafer was cleaned in an ultrasonic bath with acetone for 10 min and rinsed with isopropanol followed by baking on a hot plate at 120 °C for 180 s. Afterwards, a 2 μm

RZJ 304.50 photoresist was spin-coated on the silicon at a speed of 5000 r/s for 30 s. The substrate with photoresist was then baked on a hot plate (100 °C, 180 s). Conventional UV photolithography (SUSS-MA6) was used to transfer DSRR pattern on photoresist, and then developed with a RZX3038 developer for 26 s to 30 s. The patterned sample was then baked by a hot plate (120 °C, 90 s). Finally, electron beam evaporation (TF500) was used to deposit 200 nm thick aluminum, and liftoff of the remaining photoresist was done in bath with acetone (60 °C, 30 min).

**Measurements**

Transmission spectra were measured with commercially available terahertz time-domain spectroscopic system[33]. After Fourier transform, we obtain transmission spectra of samples ($t_s$) and references (bare silicon substrate, $t_r$), and get normalized transmission spectra ($t=t_s/t_r$). With measured transmission spectra, the total $Q_{tot}$ ($\frac{1}{Q_{tot}} = \frac{1}{Q_{rad}} + \frac{1}{Q_{ohm}}$) can be extracted by using Fano line-shape equation[34, 35]:

$$T_{Fano} = \left| a_1 + ja_2 + \frac{b}{\omega - \omega_0 + j\gamma_{tot}} \right|^2 \quad (2)$$

where $a_1$, $a_2$ and $b$ are real; $\gamma_{tot}$ is the total radiation rate; $\omega_0$ is the central frequency of resonance. $Q_{tot}$ was determined by $Q_{tot} = \omega_0 / 2\gamma_{tot}$. We could retrieve $Q_{rad}$ by estimating $Q_{Ohm}$ from simulations.

## Acknowledgements

This work was supported by the National Natural Science Foundation of China (Award No.: 62175099), and startup funding of Southern University of Science and Technology. The authors acknowledge the assistance of SUSTech Core Research Facilities and thank Yao Wang for helpful discussions on fabrication.

## Author contributions

L.C. initiated the idea and supervised the project. J. F. fabricated samples, performed experiments and simulations, analyzed data, and initiated the manuscript. J. F. and L.C. analyzed data with inputs from Z. X., H. X., D. L., G. X., J. G. and J. H. All authors wrote and commented on the manuscript.

## Competing interests

The authors declare no competing interests.

**Data Availability**

The data are available from the corresponding author upon reasonable request.